\documentclass[preprint,12pt]{elsarticle}
\usepackage{graphicx}
\usepackage{epstopdf}
\usepackage{dcolumn}
\usepackage{bm}
\usepackage{caption}
\usepackage{subfig}
\renewcommand\thesubfigure{(\alph{subfigure})}
\DeclareCaptionFormat{cont}{#1 (cont.)#2#3\par}

\usepackage{amsmath}
\begin{document}
	\title{Thermal ignition revisited with two-dimensional molecular dynamics: role of fluctuations in activated collisions}
	\author[]{N. Sirmas}
	\ead{nsirmas@uottawa.ca}
	\author[]{M. I. Radulescu}
	\ead{matei@uottawa.ca}
	\address{ Department of Mechanical Engineering, University of Ottawa, 161 Louis Pasteur, ON, K1N 6N5, Canada}
	\date{\today}
	
	\begin{abstract}
	The problem of thermal ignition in a homogeneous gas is revisited from a molecular dynamics perspective. A two-dimensional model is adopted, which assumes reactive disks of type A and B in a fixed domain that react to form type C products if an activation threshold for impact is surpassed. Such a reaction liberates kinetic energy to the product particles, representative of the heat release. The results for the ignition delay are compared with those obtained from the continuum description with the reaction rate evaluated from kinetic theory assuming local thermodynamic equilibrium and Maxwell-Boltzmann statistics, in order to assess the role played by molecular fluctuations. Ignition times obtained using molecular dynamics are ensemble averaged over 100 simulations to address the statistics of the ignition event. Results show two regimes of non-equilibrium ignition whereby ignition occurs at different times as compared to that for homogeneous ignition assuming local equilibrium. The first regime is at low activation energies, where the ignition time is found to be higher than that expected from theory for all values of heat release. The lower reaction rate is shown to occur due to a departure from local equilibrium for the different species, in agreement with predictions from Prigogine and Xhrouet.  In this low activation energy regime, the ignition times from molecular dynamics are also found to be independent of domain size and there is little variance between different realizations under similar conditions, which suggests that the ignition is spatially homogeneous. The second regime occurs at high activation energies and sufficiently large heat release values. In this high activation energy regime, ignition times are found to be dependent on domain size, where small domains of $2.87\times2.87$ mean free paths yielded longer ignition delays than predicted, while for larger domain sizes, with $9.06\times9.06$ and $28.73\times28.73$ mean free paths, shorter ignition delays than those expected were observed. Results for larger systems agree with the expectations by Prigogine and Mahieu, who demonstrate that the inclusion of a sufficiently large heat of reaction can yield a non-equilibrium reaction rate larger than expected for a homogeneous system in equilibrium. Results yield a large variance for ignition times under these conditions, which combined with the dependence on the domain size suggests a departure from homogeneous combustion.  The results obtained are in qualitative agreement with experimental observations of auto-ignition at relatively low temperatures, where hot-spot ignition and associated ignition delays lower than predicted are generally observed. 
		
	\end{abstract}
	
	\maketitle
	
	\section{Introduction}
	Ignition phenomena are central to combustion problems \cite{Williams1985}.  Experimentally, hot-spot ignition is generally observed at low temperatures \cite{meyer1971shock, gardner2005}, with ignition delays typically lower than predicted from classical chemical kinetic descriptions in a homogeneous system.  Recently, detailed experiments suggest that spark induced ignition, for example, is an intrinsically stochastic process \cite{bane2010} when observed at the continuum scale. At low temperatures, ignition phenomena are typically of the thermal type \cite{sanchez2014recent}. The present paper focuses on this type of thermal ignition in a model system.
	
	
	It has been proposed that the source of stochasticity in ignition phenomena may be attributed to fluctuations within the reactive medium, ranging from thermal fluctuations, to fluctuations in the number of activated collisions yielding ignition, hydrodynamic fluctuations from macroscopic effects \cite{borisov1974origin} and hydrodynamic instabilities. Departures from equilibrium have also been considered, especially for small-scale systems \cite{gorecki1987adiabatic, lemarchand2004enhanced}.
	
	Previous authors have demonstrated that exothermicity can play a strong role in introducing non-equilibrium effects and modifying the macroscopic rate of reactions from that obtained from the standard kinetic theory evaluation assuming local thermodynamic equilibrium (e.g., see \cite{Vincenti&Kruger1975}). Some of the earliest work by Prigogine and Xhrouet \cite{prigogine1949perturbation} estimated the change of the reaction rate by perturbation of the Maxwellian distribution for reactions with low levels of heat release, concluding that the non-equilibrium reaction rate is lower than the equilibrium rate for low activation energies. 
	
	In contrast, Prigogine and Mahieu \cite{prigogine1950perturbation} demonstrated that the inclusion of a sufficiently large heat of reaction can yield a non-equilibrium reaction rate larger than the one derived with the assumption of local equilibrium, using the same perturbation method. Similar results have been reported by others using comparable perturbation methods to a reactive system, demonstrating the roles that both activation energy and heat release in exothermic reactions have on the departure from equilibrium \cite{present1968chapman, ross1961some, shizgal1970nonequilibrium}.
	
	Microscopic models have been used to investigate the role that fluctuations and non-equilibrium effects have on reactive systems, whereby such models naturally account for fluctuations in reactive systems. Such studies have been conducted using Molecular Dynamics (MD) simulations \cite{Chou&Yip1984, Gorecki&Gorecka2000, gorecki1987adiabatic} and models using the Direct Simulation Monte Carlo (DSMC) method \cite{lemarchand2004enhanced,lemarchand2004fluctuation,mansour1992microscopic,nowakowski2002thermal, dziekan2012particle}. Recently, models involving Landau's \textit{fluctuating hydrodynamics}  formalism \cite{Landau&Lifshitz1959} that bridges the molecular and continuum descriptions have been formulated to address problems in reactive systems  \cite{bhattacharjee2015fluctuating}, although the authors argue that particle based models are still needed in order to capture physics at the molecular scale. 
	
	Despite the extent of the previous work, few studies have addressed the arguments made by Borisov \cite{borisov1974origin} regarding the role that fluctuations and hot-spot formations have on the ignition delay in auto-ignition phenomena. The present work addresses these issues by revisiting the classical problem of thermal ignition~\cite{Williams1985} via MD simulations in a simplified binary system of reactive gases. Such a system is suitable to look at low temperatures ignition, whereby the chemistry is shown to follow a thermal ignition process~\cite{sanchez2014recent}. The MD description used in the present paper addresses the potential role of spatial non-homogeneities in the ignition problem, which is typically more difficult to consider in a continuum description using corrections of the type initiated by Prigogine and co-workers~\cite{prigogine1949perturbation}, as discussed above.  
	
	The present paper reports the results obtained from microscopic simulations of the molecular dynamics using the two-dimensional hard particle method with activated reactive collisions. Such hard particle models are attractive to study because of their low computational price, while still yeilding similar results to models implementing realistic inter-particle force potentials~\cite{Vincenti&Kruger1975}. Since the pioneering work of Alder and Wainright \cite{Alder&Wainright1959}, the dynamics of hard particles can be solved by solely solving for the time of collisions among different pairs of particles, where each successive collision can be predicted analytically.  The system is evolved from collision to collision, or event to event, hence the name of the algorithm, the Event Driven Molecular Dynamics method (EDMD). The method can be readily applied to hundreds of thousands of particles on today's personal computers. In the current study, simulations are limited to 2D, as 3D simulations would require significantly more computational time using this method. It can be shown that in order to replicate a 2D system with $N_{2D}$ disks in an area of $l\times l$, approximately $N_{3D}=N_{2D}^{3/2}$ spheres would be needed in a volume with similar length scales of $l\times l\times l$ in 3D. The computational time required to advance for a specified number of collisions per particle scales with $N\log N$ using the described method~\cite{Poschel&Schwager2005}, thus requiring a 3D system to run for at least $\frac{3}{2}N_{2D}^{1/2}$ times longer than a system with similar scaling in 2D.  Despite the idealization of a system in 2D, such a simplified model using EDMD is  insightful and has been used for some of the earliest investigations of non-equilibrium reactive phenomena, such as ignition \cite{Chou&Yip1984} and detonation wave propagation \cite{Kawakatsu_etal1988}, albeit with a limited number of particles. 

	The present paper uses similar reactive dynamics assumed in these earlier papers to study the problem of homogeneous thermal ignition and compare with the predictions made from the continuum description, where the reaction rate is evaluated from classical kinetic theory arguments assuming local equilibrium \cite{Vincenti&Kruger1975}.  Calculations are done in a fixed area, where the thermal ignition problem in the continuum regime is well understood \cite{Williams1985}. In order to address the statistics of the ignition phenomena, especially at low temperatures, ensemble averaging is completed over multiple realizations for different parameters.  

	The paper is organized as follows. The first part describes the reactive system that is investigated, with the details regarding the MD simulations and the continuum description. In the second part, the results from MD are compared with those obtained from the continuum description in order to draw a conclusion of whether molecular fluctuations may be responsible for departures from homogeneous ignition and changes in the ignition delay times. 
	
	\section{Methodology}
	\subsection{Model Description}

	The present model assumes an irreversible exothermic reaction of the form
	\begin{equation}
	A+B\rightarrow C+C+heat
	\end{equation}  
	in which only collisions among the reactants A and B can yield two product species C. This reactive model is similar to that presented by previous authors  \cite{Chou&Yip1984, mansour1992microscopic, gorecki1987adiabatic, Gorecki&Gorecka2000}, with modified chemistry to allow for depletion of two distinct reactants. Such a description is desirable since fluctuations of species can be considered, and gives the flexibility of varying the ratio of reactants to study the effect of dilution on ignition. 
	
		\begin{figure}
			\centering
			\includegraphics[width=\linewidth]{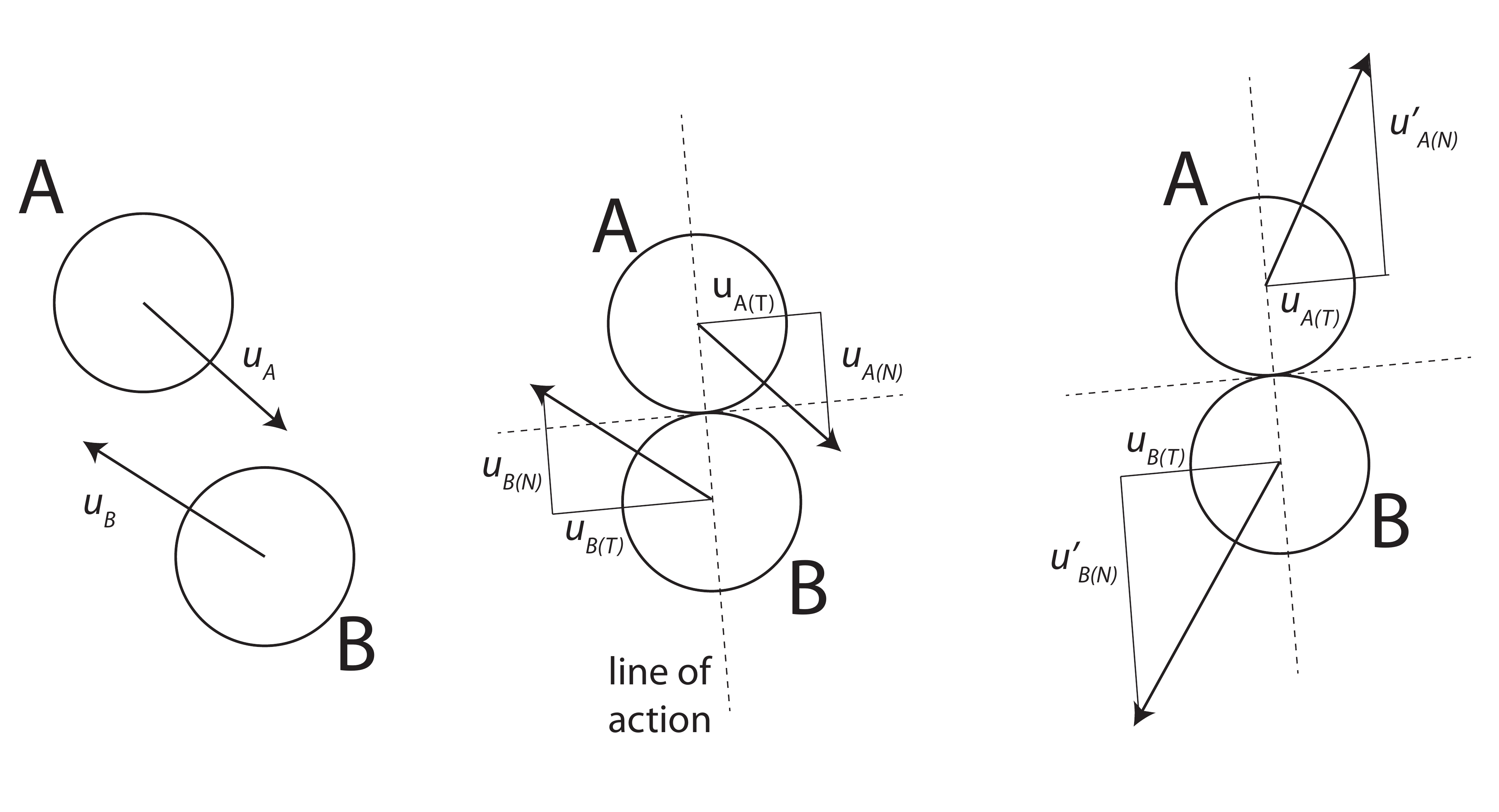}
			\caption{Kinematics of collision between particles A and B before impact(left), during impact(center) and immediately after collision(right).}
			\label{fig:Collision_model}
		\end{figure}
	
	The model assumes $N_A$ and $N_B$ number of type A and B reactive disks, respectively. Disks have a diameter $d$ occupying a prescribed volume fraction $\eta$ within a 2D fixed domain.   In the reactive collision, each existing particle A and B transform into two reacted particles C. Disks A, B, and C have identical masses. All collisions are assumed to be elastic with the exception of a reactive collision, which gives an amount of chemical energy $Q$ to each reacted particle in order to increase their kinetic energies. The changes in speeds of the reacting particles after impact occur along the line of action, while the tangential components remain unchanged. A schematic of this process is shown in Fig.~\ref{fig:Collision_model}. The activation energy necessary for reactions is taken as $E_A$. The activation energy can be related to the minimum impact velocity, $u_{cr}$, necessary for a reactive collision to occur between particles A and B,
	\begin{equation}
	\left|u_{A(N)}-u_{B(N)}\right|>u_{cr},
	\end{equation}
	which for a 2D system the relationship between $E_A$ and $u_{cr}$ is
	\begin{equation}
	u_{cr}=\sqrt{4E_A}.
	\end{equation}
	
	Video 1 in the supplementary material demonstrates the dynamics of the disks for such a system.
	
	\subsection{Molecular Dynamics Details}
	The MD model was established by implementing the collision rules for reactive and non-reactive encounters into an Event Driven Molecular Dynamics algorithm, as pioneered by Alder and Wainright~\cite{Alder&Wainright1959}, as also described by P\"{o}schel and Schwager \cite{Poschel&Schwager2005}. For each simulation, $N_A$ and $N_B$ disks of diameter $d$ are initialized in a square domain with equal speeds and randomized trajectories. Collisions with the boundaries were treated as elastic. The particles were left to thermalize prior to allowing them to undergo reactive collisions. This is shown to occur once the distribution of speeds converges to that expected from the Maxwell-Boltzmann (MB) distribution. The probability distribution function (PDF) for the MB distribution is given as
	
	\begin{equation}\label{eq:MB}
	f(v_i)=\frac{mv_i}{2RT}\exp\left(-\frac{mv_i^2}{2RT}\right),
	\end{equation}
	with a corresponding cumulative distribution function (CDF) of
	\begin{equation}\label{eq:CDF}
	F(v_i)=\int_{0}^{\infty}f(v_i)dv_i
	\end{equation}

	\begin{figure}
	\centering
	\captionsetup[subfigure]{oneside,margin={1cm,.5cm}} 
	\subfloat[$t=\tau_o$]{\includegraphics[width=0.45\linewidth]{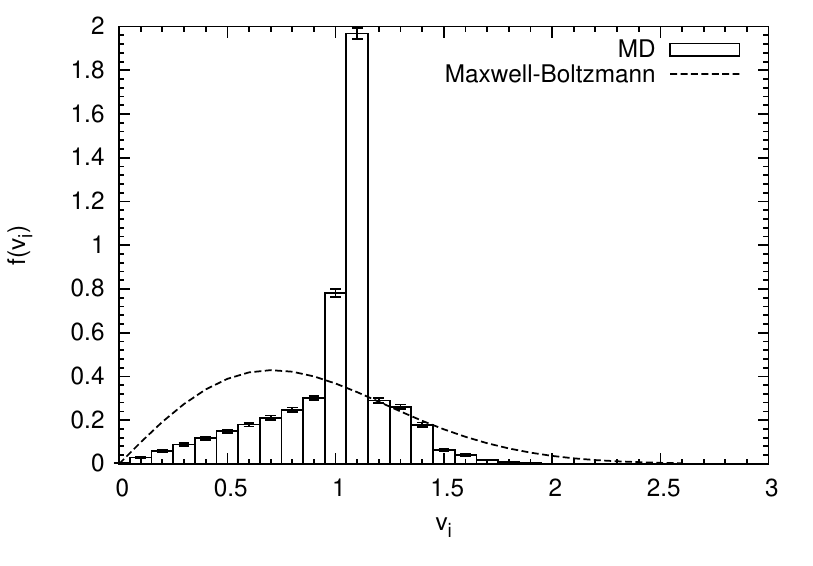}}~
	\subfloat[$t=5\tau_o$]{\includegraphics[width=0.45\linewidth]{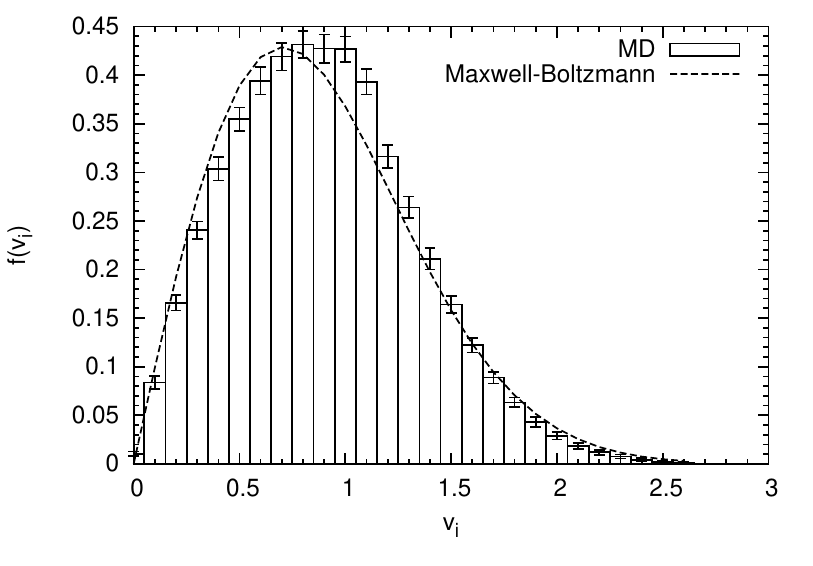}}\\
	\subfloat[$t=10\tau_o$]{\includegraphics[width=0.45\linewidth]{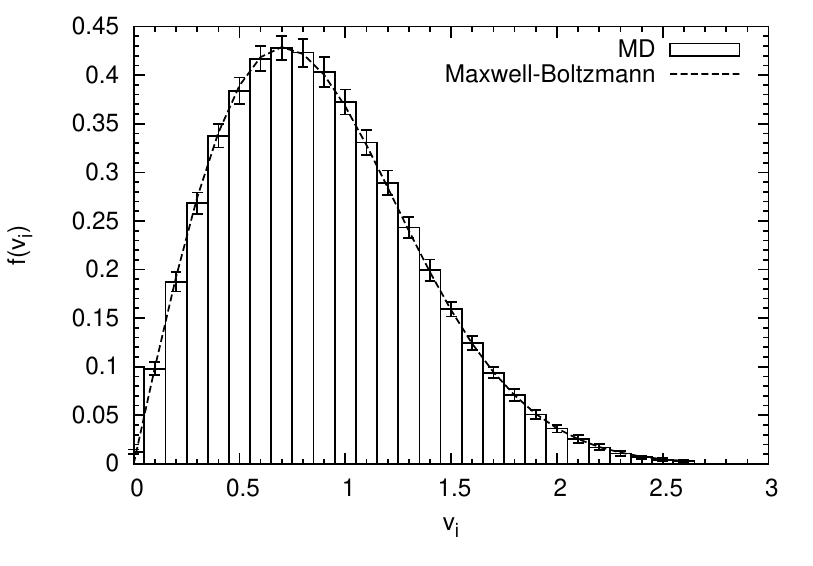}}~
	\subfloat[$t=50\tau_o$]{\includegraphics[width=0.45\linewidth]{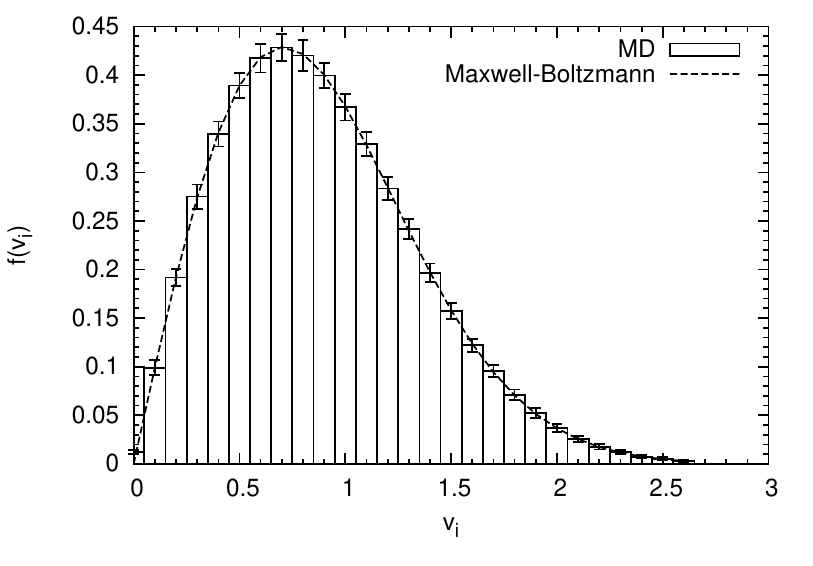}}
	\caption{Probability distribution of speeds obtained from MD for different points in time for $N=10000$ disks and ensemble averaged over 100 simulations, as compared to that expected from Maxwell-Boltzmann statistics, with error bars represent the standard deviation given from the ensemble. Speeds are normalized by the initial speed of the disks and time is given in terms of the mean collision time $\tau$. }
	\label{fig:MB_thermalize}
	\end{figure}
	
An example of the thermalization process is given in Fig.~\ref{fig:MB_thermalize}, shown with the PDF of the distribution of speeds ensemble averaged over 100 simulations for a system of $N=10000$ disks, as compared with the MB distribution at different times. The error bars represent the standard deviation of the ensemble average for a particular speed. After one mean collision time, shown in Fig.~\ref{fig:MB_thermalize}(a), there is a large peak for the distributions at the initial speed of the disks. After five mean collision times, shown in Fig.~\ref{fig:MB_thermalize}(b), the distribution approaches the MB distribution, although there is a plateau that still occurs at the initial speed of the disks. At ten mean collision times, shown in Fig.~\ref{fig:MB_thermalize}(c), the distribution of speeds converges to the MB distribution, and does not change over time, as shown by the distribution after fifty mean collision times in Fig.~\ref{fig:MB_thermalize}(d). This shows that the system has come to equilibrium after ten mean collision times, and this is found to be a sufficient time to equilibrate the system. Similar results are found for different number of disks in the system, with increasing noise for smaller systems.

Reactive collisions are allowed after this initial thermalization process over ten mean collision times, which marks time zero. The evolution of temperature and species for the entire domain was output at prescribed time intervals.  Validation of the numerical implementation can be found elsewhere \cite{Sirmasetal2012, Sirmas2015}.

\subsection{Homogeneous Ignition at the Continuum Level}
The results obtained using the molecular dynamic description detailed above were compared with those obtained from a continuum description assuming a homogeneous system.  For a constant volume evolution, only the evolution of the system's partition of energy and composition applies~\cite{Williams1985}: 
	\begin{equation}
	\rho c_V\frac{dT}{dt}=Q\omega_C
	\end{equation}
	\begin{equation}
	\rho\frac{dY_C}{dt}=\omega_C
	\end{equation}
where $Y_C$ and $\omega_C$ are the mass fraction of products C and their production rate, respectively.  The reaction rate $\omega_C$ for activated reactions can be obtained from classic kinetic theory arguments \cite{Vincenti&Kruger1975} if one makes the standard assumption of local Maxwell-Boltzmann equilibrium and a homogeneous system.  For the two-dimensional system of the present study, it takes the form~\cite{sirmas2014evolution}:
	\begin{equation}\label{eq:rate}
	\omega_C=\frac{16}{d\sqrt{\pi}}g_2(\eta)\eta\rho Y_AY_B\sqrt{RT}\exp\left(-\frac{E_A}{RT}\right)
	\end{equation}
where $Y_A$, and $Y_B$ are the mass fractions of reactants A and B, respectively, with the pair correlation function 
	\begin{equation}
	g_2(\eta)=\frac{1-\frac{7}{16}\eta}{(1-\eta)^2}.
	\end{equation}
The described system is characterized by five parameters, the volume fraction $\eta$, the specific heat $c_v=R$ for a 2D hard particle system, the initial temperature $T_o$ and the parameters $Q$ and $E_A$.  When non-dimensionalized, the parameters $\eta$, $Q/RT_0$ and $E_A/RT_0$ uniquely characterize the system's evolution.
\subsection{Parameters and Scaling}
The current study considers a system in the dilute, ideal gas regime, with a volume fraction of $\eta=0.01$. The length scales are normalized by the mean free path in the initial gas,
	\begin{equation}
	\lambda=\frac{d\sqrt{\pi}}{4\sqrt{2}g_2(\eta)\eta}.
	\end{equation}
The time scales are normalized by the initial mean free time, with $\tau_o=\lambda/u_{rms(o)}$. With such scaling, the homogeneous ignition description becomes independent of $\eta$.  
	
	The effect of the domain size is also considered in the molecular dynamic calculations.  This is achieved by investigating systems with different numbers of particles, while maintaining a constant volume fraction $\eta$. A summary of the conditions considered are shown in Table \ref{table:params}.
	\begin{table} 
		\begin{center}
			\caption{Parameters considered for molecular dynamics simulations}\label{table:params}
			\begin{tabular}{ccc}
				\hline Total number of particles & Particle diameter & Domain size \\ 
				$N$ & $d/\lambda$  & $(L_x\times L_y)/\lambda$\\
				\hline 100 & 0.0324 & 2.87 $\times$ 2.87 \\ 
				1000 & 0.0324 & 9.06 $\times$ 9.06 \\ 
				10000 & 0.0324 & 28.73 $\times$ 28.73 
			\end{tabular}
		\end{center}
	\end{table}

	\section{Results and discussion}
	
\begin{figure}
		\centering
		\includegraphics[width=0.8\linewidth]{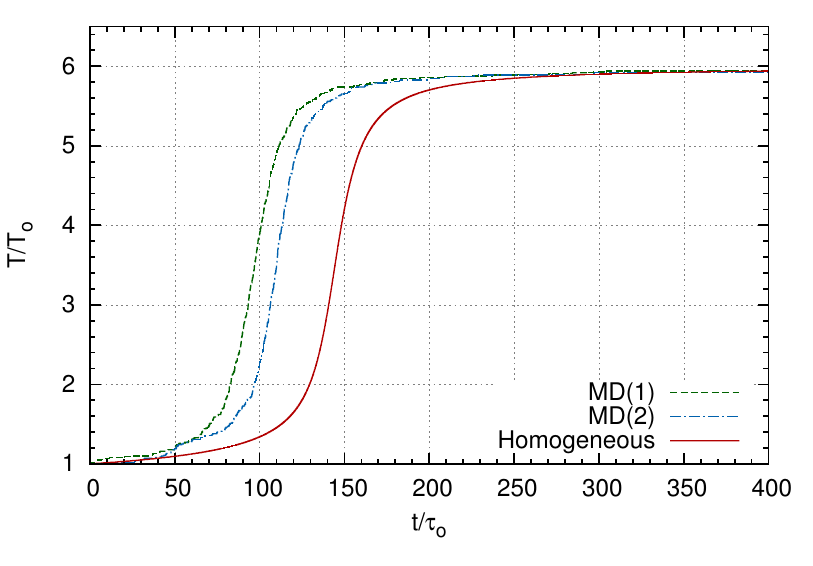}
		\caption{Example evolution of temperature obtained from MD  and for homogeneous ignition with $Q/RT_o=5$ and $E_A/RT_o=7.5$, for two separate realizations from MD with 1000 disks.}
		\label{fig:Q5Ea7_5_example}
	\end{figure}

	An example of the evolution of temperature obtained from the continuum model and from two separate realizations in MD is shown in Fig.~\ref{fig:Q5Ea7_5_example} for the case of $Q/RT_o=5$ and $E_A/RT_o=7.5$, with $N_A=N_B=500$ particles. While all three curves display the characteristics of thermal ignition, marked by a slow induction phase followed by a rapid thermal run-away \cite{Williams1985}, it is evident that the molecular dynamic results can have substantial differences from the continuum prediction.  Furthermore, different realizations, obtained by different randomized trajectories of the initialized particles, also yields a different outcome owing to the molecular noise.   
	
	In the subsequent discussion, the ignition delay time in a single realization is defined as the time elapsed until 50\% of the least abundant reactant is depleted.  The mean ignition delay time $t_{ig}$ from the MD calculations was obtained by averaging the ensemble of single realization ignition delay times obtained for 100 simulations. The ignition delay obtained from the continuum model assuming a homogeneous mixture with negligible fluctuations is defined as $t_{ig,h}$.  
	
\begin{figure}
		\centering\includegraphics[width=0.8\linewidth]{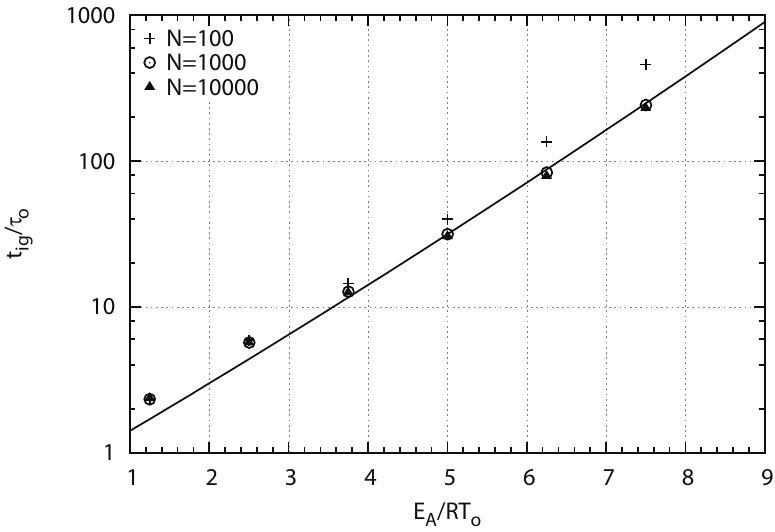}
		\caption{Comparison of ignition times obtained from MD for $N_A=9N_B$ (data points, averaged over 100 realizations) with that obtained for homogeneous ignition (solid line), for $Q/RT_o=7.5 $.}
		\label{fig:Q7_5_diffconcentrations}
	\end{figure}

The first case considered is that where the number of one reactant far exceeds the other, i.e., the diluted case. Figure \ref{fig:Q7_5_diffconcentrations} shows the Arrhenius plot for ignition delay time for the mixture $9A + B$, for different domain sizes with $Q/RT_o=7.5$ and varying $E_A$. At low activation energies, the results for $t_{ig}$ are the same for all domain sizes of $N=100$, 1000, and 10000, yielding a higher value than $t_{ig,h}$. This deviation will be discussed further in the next section. In contrast, for high activation energies and sufficiently large domains, with totals of 1000 and 10000 particles, results for $t_{ig}$ agree well with $t_{ig,h}$. However, for the smaller domain of 100 particles, the domain is too small to capture the correct ignition delay, giving higher than expected values.

	\begin{figure}
		\centering\includegraphics[width=0.8\linewidth]{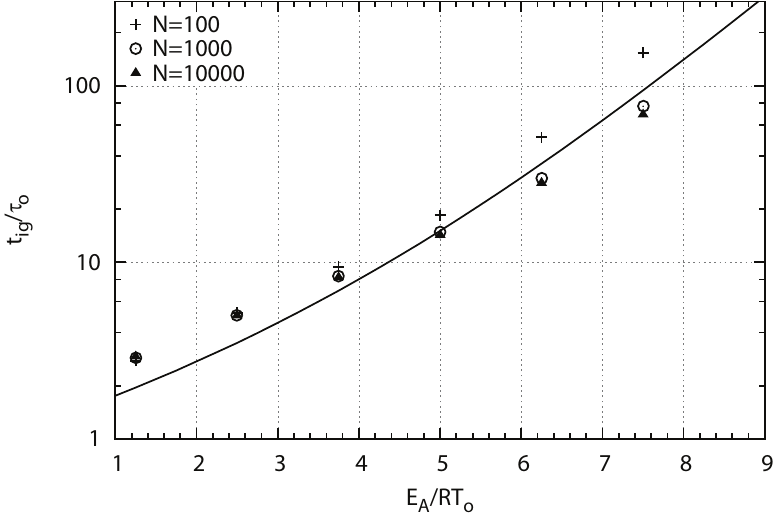}
		\caption{Comparison of ignition times $t_{ig}$ normalized my the initial mean free time obtained from MD for different number of disks $N$, where $N_A=N_B=N/2$ (data points, averaged over 100 realizations) with that obtained for homogeneous ignition (solid line), for $Q/RT_o=7.5 $.}
		\label{fig:Q7_5_diffN}
	\end{figure}

The non-diluted case is however more interesting. Figure~\ref{fig:Q7_5_diffN} shows the results obtained for the different domain sizes with $Q/RT_o=7.5$ and varying $E_A$ for equal parts of reactants A and B. Results show clear differences between the ignition times expected from homogeneous ignition assuming local equilibrium and those obtained via MD. 

At low activation energies, MD results for all domain sizes give an ignition time greater than for homogeneous ignition, similar to what was observed for the diluted case above. At these activation energies, the results show that $t_{ig}$ is independent on the domain size. However, results show a dependency on the domain size as the activation energy is increased. For a small domain size, with $N= 100$, $t_{ig}$ from MD is larger than $t_{ig,h}$ for all values of activation energy. For larger domain sizes, with $N=1000$ and 10000, the ignition time obtained from MD exceeds the estimate from the continuum model for smaller $E_A$, but becomes lower than the continuum prediction for larger values of $E_A$. 

\begin{figure}
	\centering
	\captionsetup[subfigure]{ oneside,margin={1cm,.5cm}} 
	\subfloat[$Q/RT_o=2.5$]{\includegraphics[width=0.8\linewidth]{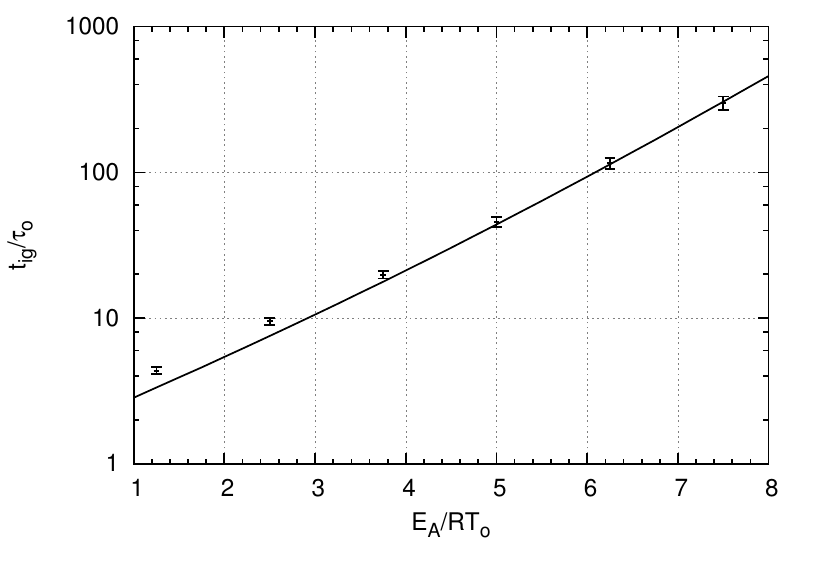}}\\\vspace*{-10pt}
	\subfloat[$Q/RT_o=5$]{\includegraphics[width=0.8\linewidth]{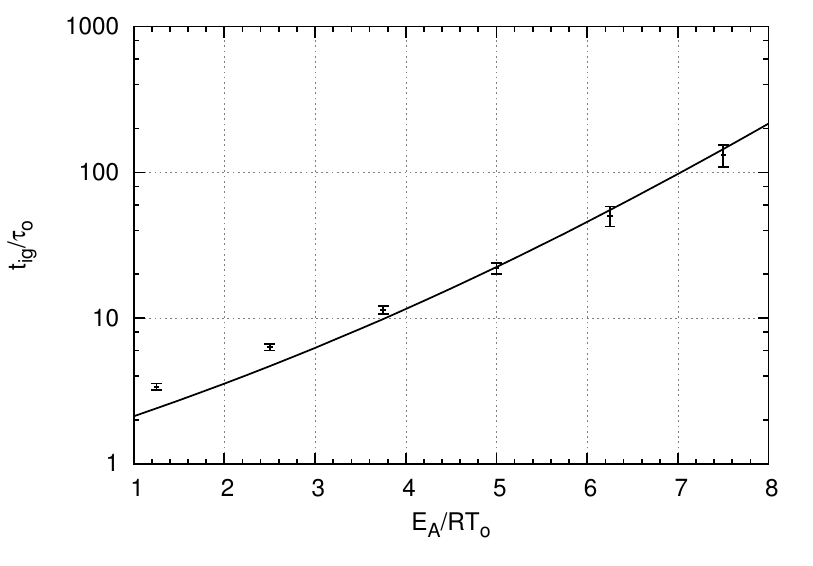}}
	
	\caption{Ignition times obtained for homogeneous ignition (solid line) compared with results from MD for varying $Q$ obtained with $N_A=500$ disks and $N_B=500$ disks (data points, averaged over 100 realizations) with error bars representing the standard deviation from ensemble averaging.}
	\label{fig:compare_MDv0D}
\end{figure}
\renewcommand{\thefigure}{\arabic{figure} (Cont.)}
\addtocounter{figure}{-1}
\begin{figure}
	\centering
	\addtocounter{subfigure}{+2}
	\captionsetup[subfigure]{ oneside,margin={1cm,.5cm}} 
	\subfloat[$Q/RT_o=7.5$]{\includegraphics[width=0.8\linewidth]{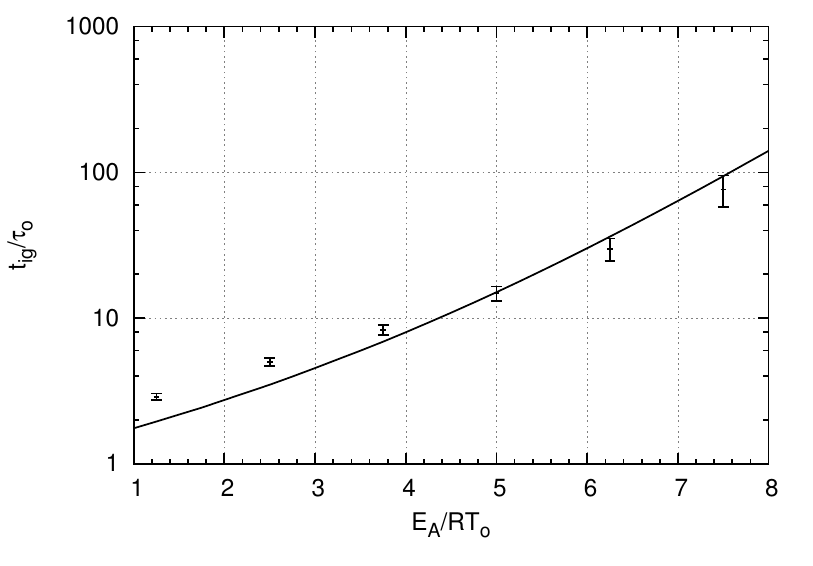}}\\\vspace*{-10pt}
	\subfloat[$Q/RT_o=10$]{\includegraphics[width=0.8\linewidth]{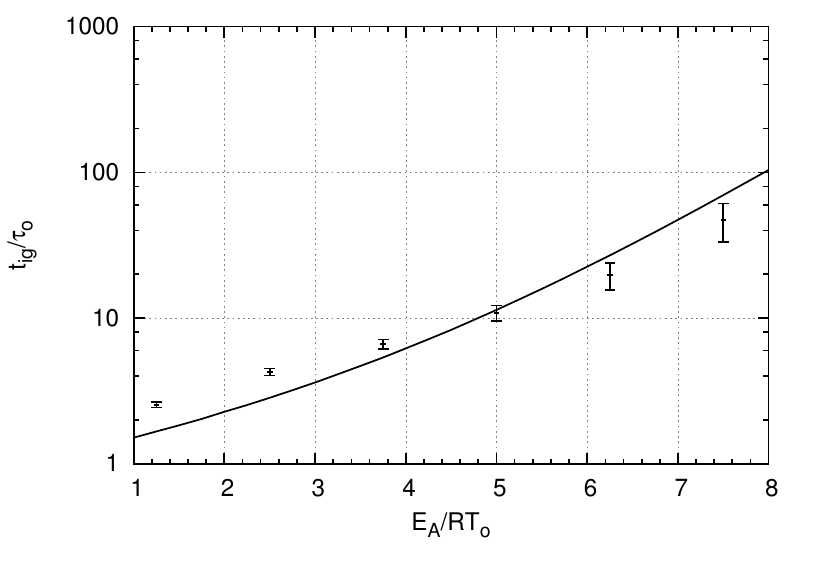}}
	\caption{Ignition times obtained for homogeneous ignition (solid line) compared with results from MD for varying $Q$ obtained with $N_A=500$ disks and $N_B=500$ disks (data points, averaged over 100 realizations) with error bars representing the standard deviation from ensemble averaging.}
	\label{fig:compare_MDv0D_cont}
\end{figure}
\renewcommand{\thefigure}{\arabic{figure}}

\begin{figure}
		\centering
		\captionsetup[subfigure]{ oneside,margin={1cm,.5cm}} 
		\subfloat[]{\includegraphics[width=0.8\linewidth]{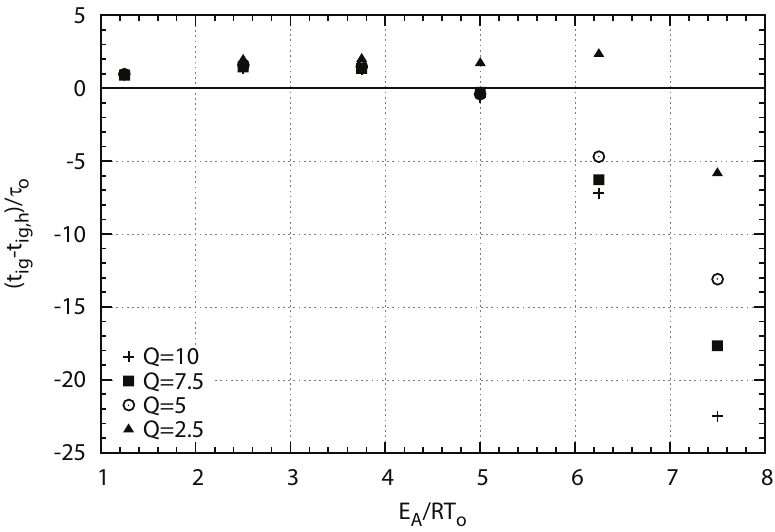}}\\\vspace*{-10pt}
		\subfloat[]{\includegraphics[width=0.8\linewidth]{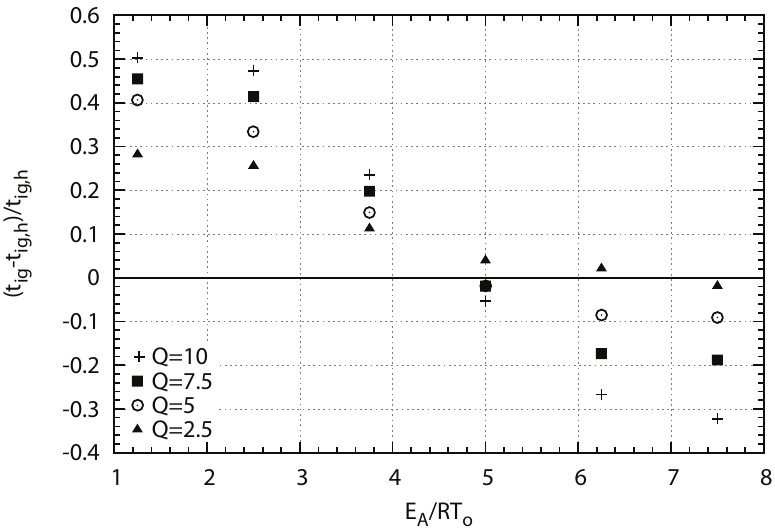}}
		\caption{(a) Difference and (b) relative difference between the ignition times obtained from MD and those calculated by assumed a homogeneous ignition in local equilibrium, for the cases presented in Figure~\ref{fig:compare_MDv0D}.}
		\label{fig:error}
	\end{figure}

Figures~\ref{fig:compare_MDv0D}(a)-(d) show how $t_{ig}$ compares with $t_{ig,h}$ for varying values of $Q$ and $E_A$, for $N=1000$. The difference and relative difference between $t_{ig}$ and $t_{ig,h}$ are shown for these cases in Figs.~\ref{fig:error}(a)-(b). 
	
At low activation energies, $t_{ig}$ was found to be larger than $t_{ig,h}$ for all heat release parameters investigated.  In Fig.~\ref{fig:error}(a) it can be seen that for $E_A/RT_o=1.25-3.75$ the difference between $t_{ig}$ and $t_{ig,h}$ is independent of heat release, with MD yielding a longer ignition delay by approximately 1-3$\tau_o$. This longer ignition delay can be attributed to the non-equilibrium effects for the rapid reactions. 

\begin{figure}
		\centering
		\captionsetup[subfigure]{ oneside,margin={1cm,.5cm}} 
		\subfloat[$E_A/RT_o=2.5$, $Q/RT_o=2.5$]{\includegraphics[width=0.5\linewidth]{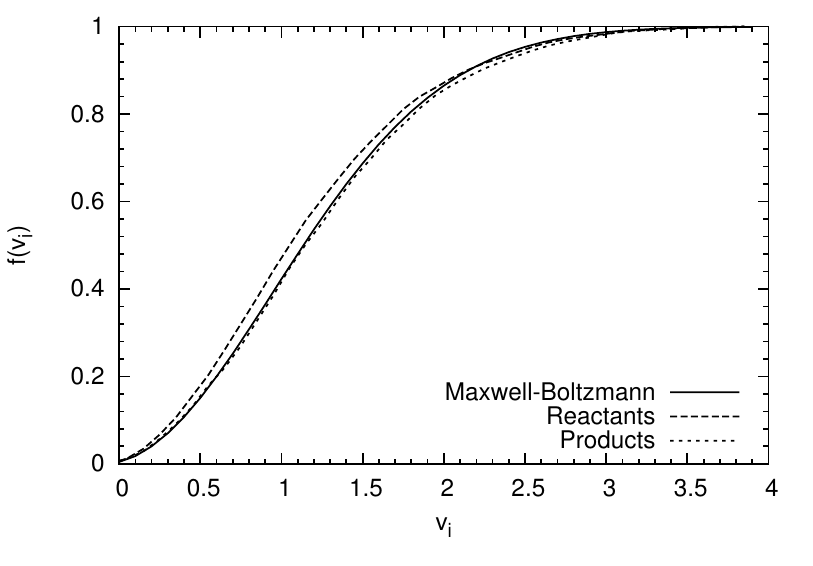}}	
		\subfloat[$E_A/RT_o$=2.5, $Q/RT_o=10$]{\includegraphics[width=0.5\linewidth]{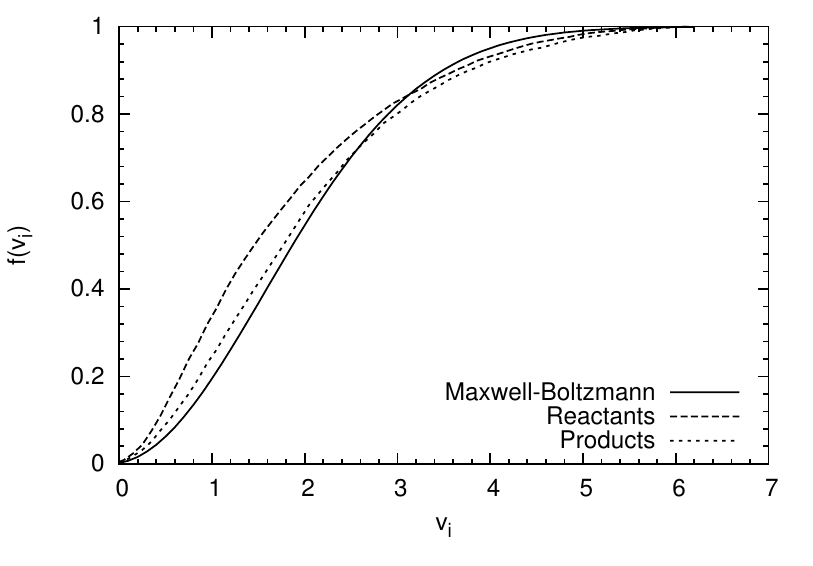}}	
	\\\vspace*{10pt}
		\subfloat[$E_A/RT_o=7.5$, $Q/RT_o=2.5$]{\includegraphics[width=0.5\linewidth]{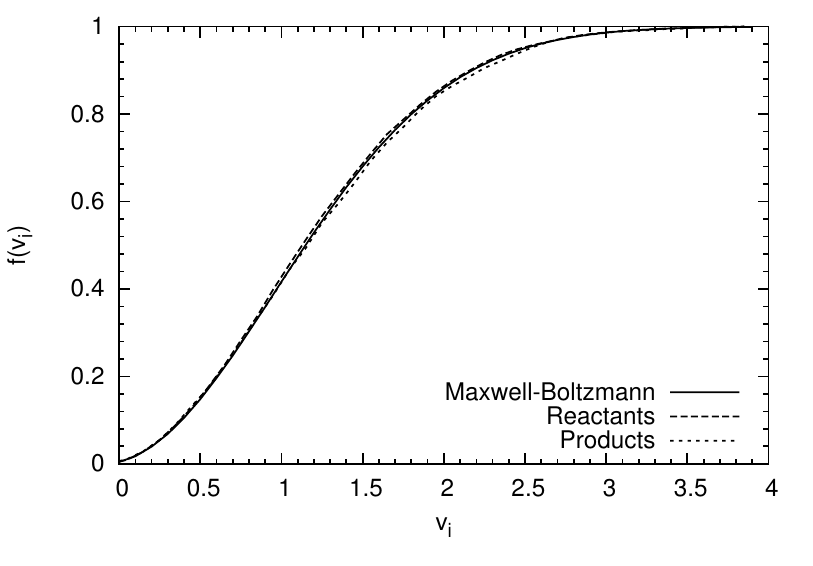}}	
		\subfloat[$E_A/RT_o=7.5$, $Q/RT_o=10$]{\includegraphics[width=0.5\linewidth]{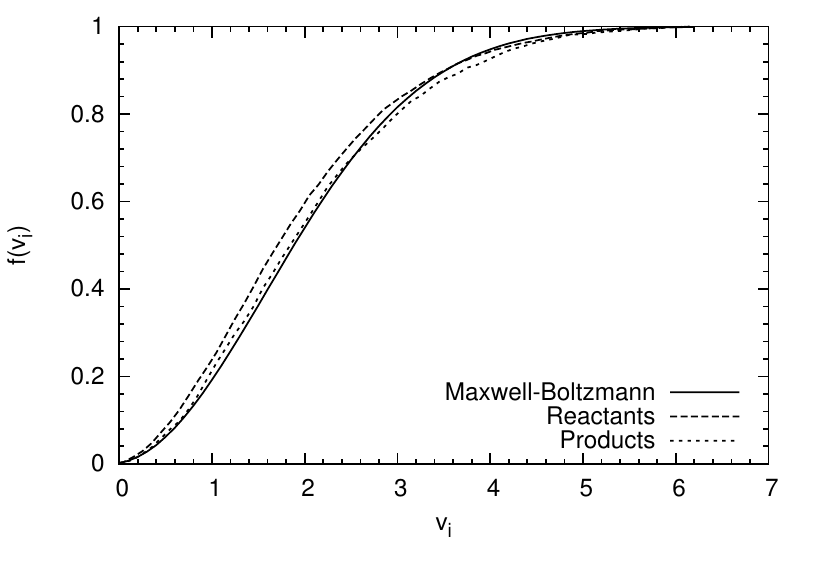}}
		\caption{Cumulative distribution functions for reactants (disks A and B) and products (disks C) obtained from MD at the time of ignition compared to that expected for a MB distribution, for different values of activation energy and heat release with $N=10000$. Speeds are normalized by the initial speed of the disks $u_{rms(o)}$.}
		\label{fig:CDF compare}
	\end{figure}

Figure \ref{fig:CDF compare} shows how the CDF for speeds of reactants (disks A and B) and products (disks C) compare to the MB distribution at $t_{ig}$ for $N=10000$ disks. For $E_A/RT_o=2.5$ and $Q/RT_o=2.5$, shown in Fig.~\ref{fig:CDF compare}(a), reactants are shown to depart from local equilibrium, shown by the shifting of the CDF to the lower speeds. As the heat release is increased to $Q/RT_o=10.0$, with an activation energy of $E_A/RT_o=2.5$, shown in Fig.~\ref{fig:CDF compare}(b), a larger departure from equilibrium is seen with the CDF for reactants and products differing from MB. This finding confirms the predictions of Prigogine and Xhrouet~\cite{prigogine1949perturbation} whereby the reaction rate for low activation energy is slower than that expected for a system in equilibrium due to a departure from equilibrium for the different species, even in the absence of any heat release.

 Figure~\ref{fig:error}(a) and (b) show that the heat release has a stronger effect on the difference between $t_{ig}$ and $t_{ig,h}$ as the activation energy increases further. At $E_A/RT_o=5.0$, results for $Q/RT_o=5, 7.5$ and 10 show a transition whereby $t_{ig}$ is lower than $t_{ig,h}$. This transition occurs between $E_A/RT_o=6.25$ and 7.5 for $Q/RT_o=2.5$. As $E_A$ increases after these transitions, larger values of $Q$ give a larger difference between $t_{ig}$ and $t_{ig,h}$. This can clearly be seen at $E_A/RT_o=7.5$ where $Q/RT_o=2.5$ yields $t_{ig}$ being approximately $5\tau_o$ less than $t_{ig,h}$, representing a relative difference of approximately 1\% with $t_{ig,h}$. The CDF for this case, shown in Fig.~\ref{fig:CDF compare}(c) shows that the reactants and products collapse near the MB distribution, representing a near-equilibrium ignition. In contrast, results for $Q/RT_o=10$ gives a $t_{ig}$ approximately $23\tau_o$ less than $t_{ig,h}$, yielding a relative difference of approximately 30\% with $t_{ig,h}$. The CDF for this case is shown in Fig.~\ref{fig:CDF compare}(d) where the distributions for reactants and products depart from the MB distribution.
  
 These results show that a higher heat release compared to activation energy favours a greater departure from an equilibrium homogeneous ignition, and thus yield a larger reaction rate. The departure from equilibrium at these high activation energy and levels of heat release agree well with the findings by Prigogine and Mahieu \cite{prigogine1950perturbation}, who showed that an increasing heat of reaction can yield a reaction rate larger than the one derived with the assumption of local equilibrium, which yields a shorter ignition time.

\begin{figure}
		\centering
		\includegraphics[width=0.8\linewidth]{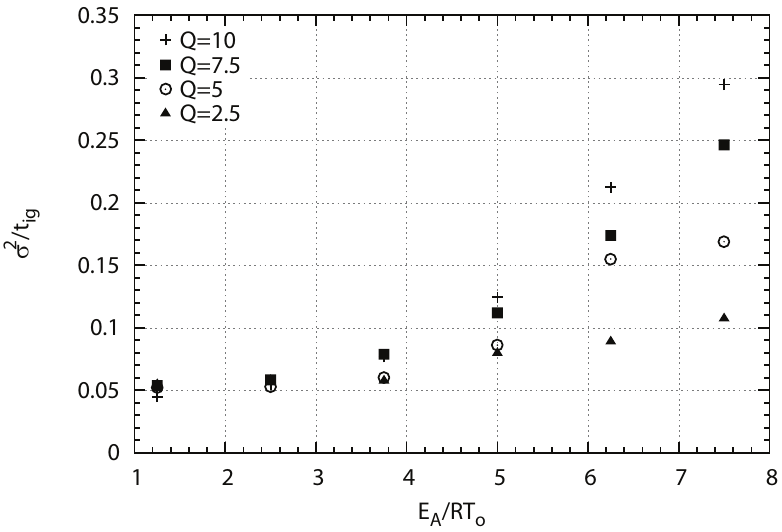}
		\caption{Relationship between standard deviation of ignition time obtained via MD compared to $E_A$ for $N=1000$ with varying $Q$.}
		\label{fig:var2}
	\end{figure}

The example evolution shown in Fig.~\ref{fig:Q5Ea7_5_example} shows that no single MD simulation will be identical, which can be attributed to the statistical fluctuations within the systems. In order to quantify the differences, the standard deviation for the ensemble averaged ignition time was calculated over the 100 simulations. The error bars shown in Fig.~\ref{fig:compare_MDv0D} for each of the data points represent the standard deviation over all simulations. Figure \ref{fig:var2} shows the standard deviation normalized by $t_{ig}$ for the cases presented in Fig.~\ref{fig:compare_MDv0D}. 
	
Results show that for a given $Q$, the standard deviation increases with increasing $E_A$. At low values of $E_A$, the standard deviation is not influenced by the heat release significantly, with all values of $Q$ yielding a standard deviation of about 5\% of the $t_{ig}$. However, for increasing values of $E_A$, the standard deviation increases with $Q$. This is shown to be substantial, up to 30\% of $t_{ig}$, as seen for $E_A/RT_o=7.5$ and $Q/RT_o=10.0$.  These results show that statistical fluctuations become important with increasing $Q$ and $E_A$, yielding less predictable ignition times.
	
Results from ensemble averaging for different levels of heat release and activation energies reveal two different non-equilibrium regimes. The first regime is at low activation energies. In this regime, results for the ignition delays are the same regardless of the system size, with little variance between simulations. The decoupling of the ignition delays from the system size suggests that ignition events under these conditions are homogeneous. The departure from the equilibrium reaction rate can be attributed to the reactions occurring at a faster rate than they are able to transfer their kinetic energy to the non-reacted disks, which is seen by the CDF of speeds for the reactants at low activation energies being shifted to the lower speeds as compared to the products, as in Fig.~\ref{fig:CDF compare}(a) and (b). It is these non-equilibrium effects that yield a higher than expected ignition time. 

The second regime is at high activation energies (i.e., the low temperature regime) and sufficiently high levels of heat release. At higher activation energies it is found that the ignition event becomes dependent on the system size, with a large variance seen between simulations under identical parameters. This behaviour suggests that spatial effects have a strong influence on ignition, characterizing such a phenomena as a non-homogeneous event. The non-homogeneity may be attributed to statistical fluctuations in particle densities and velocities, yielding sections that are more probable to react, acting as a seed for further reactions within the system. This regime has previously been identified by Prigogine and Mahieu~\cite{prigogine1950perturbation} to yield a lower than expected ignition time due to non-equilibrium effects, which is confirmed by the average ignition times obtained here. However, the present results also show that the fluctuations at the microscopic level can alter the non-equilibrium reaction rates even further, as displayed by the large variance in ignition times. 

The different regimes can also be observed at the microscopic level, by tracking how the product species form at the initial stages. Example evolutions of particle position and type are shown for single realizations in Video 2 of the supplementary material for $Q/RT_o=2.5$, $E_A/RT_o=2.5$, and Video 3 of the supplementary material for $Q/RT_o=10.0$, $E_A/RT_o=7.5$, where both examples have $N=1000$ disks. The appearance of products is shown in black for these examples. Also shown is the evolution of temperature for the simulations compared to that expected for a homogeneous system assuming local equilibrium.

For the case shown in Video 2, product species appear throughout the system as ignition commences. This evolution shows that the initial reactions are not directly influenced by previous reactions as there is not sufficient amount of time to travel and distribute kinetic energy throughout the domain. Such an ignition shows the homogeneity of ignition during the early stages. In contrast, the case shown in Video 3 for high activation energies shows how the initial reactions strongly influence subsequent reactions. This is seen by reactions forming in the proximity of product species. The non-homogeneity is evident for this case, as there are only a few distinct reactions at early times which are the seeds for further reactions and eventual thermal runaway.

The results of the present investigation suggest that the thermal ignition of gases is prone to a decrease of the ignition delays in the limit of high activation energy, as compared with the continuum prediction.  In this limit, the variability of ignition delay also increases.  This result was rationalized as being due to the increasing role in the fluctuations and the development of hot-spots.  This is in general good accord with experimental observations of shock tube ignition phenomena \cite{meyer1971shock}, where two distinct ignition regimes have been observed. At high temperatures, the ignition is homogeneous, and called ``strong".  At sufficiently low temperatures, the ignition originates from discrete centers, and is called ``mild".  In the past, the transition between the two regimes has been suggested to be governed by a unique coherence parameter $\chi$ and it is worthwhile comparing the current results with these previous findings \cite{meyer1971shock, radulescu2003propagation, radulescu2013universal}.  This parameter was suggested to control the regimes of auto-ignition in shock induced ignition, detonations, deflagration to detonation transition and engine knock.

\begin{figure}
\centering
\includegraphics[width=\linewidth]{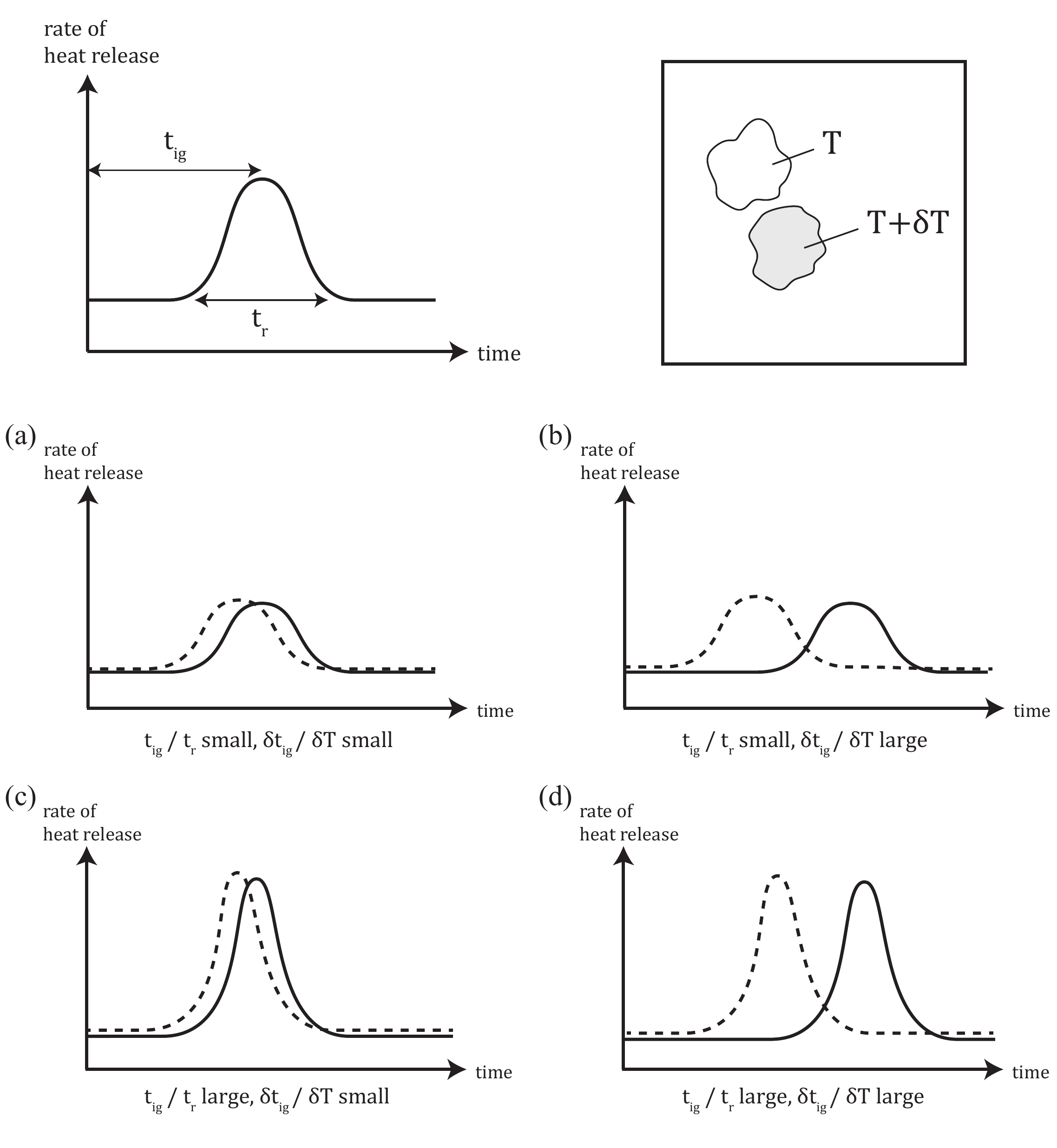}
\caption{Illustration of the coherence concept between neighbouring power pulses, given by the exothermicity profiles for two neighbouring gas elements at an initial shock state differing by $\delta T$: a) small activation energy and small
relative exothermicity, b) large activation energy, small exothermicity, c) small activation energy, large exothermicity, and d) large activation energy and large exothermicity.}
\label{fig:coherence}
\end{figure}

\begin{figure}
\centering
\includegraphics[width=0.7\linewidth]{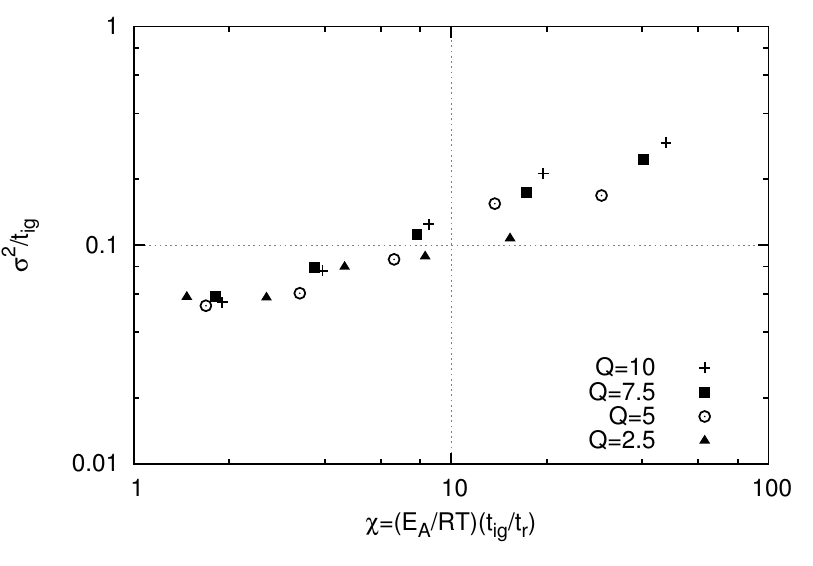}
\caption{Relationship between standard deviation in ignition time obtained via MD for a system of 1000 disks compared to the coherence parameter $\chi$ for varying Q.}
\label{fig:var2_chi}
\end{figure}

Consider two discrete zones in a homogeneous system of slightly different temperatures.  Homogeneous ignition is expected to prevail if power pulses from each zone are sufficiently coherent in time and can essentially overlap.   This will happen when the sensitivity of the ignition delay to temperature fluctuations is sufficiently low, as shown schematically in Fig.~\ref{fig:coherence}.  The decoherence will be accentuated when the reaction time is short, as shown schematically in Fig.~\ref{fig:coherence}.  Thus, as first pointed out by Soloukhin in a comment to Meyer and Oppenheim’s work \cite{meyer1971shock}, stability is more accurately described by the sensitivity to
temperature fluctuations of the characteristic induction time relative to the characteristic exothermic reaction times.  Given that the ignition delay $t_{ig}$ is proportional to $\exp (E_A/RT)$, the stability criterion becomes

\begin{equation}
\chi= \frac{T}{t_r}\frac{d t_{ig}}{dT}=\frac{t_{ig}}{t_r}\frac{E_A}{RT}
\end{equation}

\noindent where $t_r$ is the characteristic reaction time, defined in the current study as $\omega ^{-1}_{C,max}$.  Figure \ref{fig:var2_chi} shows the effect of this ignition coherence parameter on the variability of the ignition time obtained from MD. For this case the $\chi$ is obtained from the continuum model for a specified $E_A$ and $Q$. The results show the pronounced increase in the stochasticity of ignition delay with the coherence parameter $\chi$, independent of the heat release (as it directly effects the reaction time in a thermal explosion).  Analysis of the available experiments conducted by Meyer and Oppenheim \cite{meyer1971shock} reveals that typical values of $\chi$ that are approximately orders of magnitude larger lead to spotty ignition.  While the present results indicate that the correct experimental trends have been recovered \cite{Wang}, the large differences between the values of $\chi$ in the experiments and those relevant to a thermal explosion for a single reaction of the present study indicate the necessity of modelling the complex network of reactions and their sensitivity to molecular and macroscopic fluctuations.   

	\section{Conclusion}\label{sec:conclusion}
	
The problem of thermal ignition was investigated in a model 2D molecular system to address the role played by fluctuations on ignition delay and homogeneity of the process. For systems with low activation energies, the thermal ignition process is shown to be a homogeneous event with a departure from local equilibrium, contributing to a smaller reaction rate as compared to that predicted at the continuum level assuming local equilibrium.  For sufficiently large activation energies and heat release parameters, the ignition delays obtained from the molecular dynamic simulations are found to be significantly shorter than predicted at the continuum level. Analyzing the ensemble average of the simulations, it is found that fluctuations between different realizations increase with the activation energy, suggesting that the thermal ignition phenomenon is a genuine stochastic process when observed at the continuum level. These results suggest departures from the homogeneous assumption as activation energy increases, demonstrating that ignition may be due to hot-spot formations at high activation energies. The results obtained are in good agreement with experimental observations of auto-ignition at relatively low temperatures, where hot-spot ignition and associated lower ignition delays than predicted are generally observed. The findings from this paper indicate the need for further investigations to obtain quantitative correction factors, applicable to the individual rates controlling an ignition process. These studies would need to overcome the idealizations of the current paper by investigating the dynamics of individual reactions in three dimensions in the presence of many simultaneous reactions.

	\section*{Acknowledgements}
	N.S acknowledges funding through the Alexander Graham Bell Canada Graduate Scholarship (NSERC) and an Ontario Graduate Scholarship.  M.I.R wishes to acknowledge financial support from NSERC via a Discovery Grant entitled ``Predictability of detonation wave phenomena: influence of diffusive processes, hydrodynamic instabilities and relaxation effects on the hydrodynamic description".
	\bibliographystyle{elsarticle-num} 
	\bibliography{references}
	
\end{document}